\documentclass[aps,prl,preprint, titlepage,showpacs]{revtex4}

\usepackage{graphicx}

\usepackage{dcolumn}

\usepackage{bm}

\begin{document}

\title{Superdiffusion and non-Gaussian statistics in a driven-dissipative 2D dusty plasma}

\author{Bin Liu}
\author{J. Goree}
\affiliation{Department of Physics and Astronomy, The University
of Iowa, Iowa City, Iowa 52242}

\date{\today}

\begin{abstract}

Anomalous diffusion and non-Gaussian statistics are detected
experimentally in a two-dimensional driven-dissipative system. A
single-layer dusty plasma suspension with a Yukawa interaction and
frictional dissipation is heated with laser radiation pressure to
yield a structure with liquid ordering. Analyzing the time series
for mean-square displacement (MSD), superdiffusion is detected at
a low but statistically significant level over a wide range of
temperature. The probability distribution function (PDF) fits a
Tsallis distribution, yielding $q$, a measure of non-extensivity
for non-Gaussian statistics.

\end{abstract}

\narrowtext

\maketitle

Two-dimensional (2D) physical systems with a single-layer of
mutually interacting particles include a Wigner lattice of
electrons on a liquid helium surface~\cite{Grimes:79}, ions
confined magnetically in a Penning trap~\cite{Mitchell:99},
colloidal suspensions~\cite{Murray:90}, vortex arrays in the mixed
state of type-II superconductors~\cite{Gammel:87}, and dusty
plasmas levitated in a single layer~\cite{Nosenko:2004}.  Within
biological cell membranes, proteins can undergo 2D
diffusion~\cite{Jacobson:1995}.


We investigate random motion in driven-dissipative 2D liquids. In
experiments with a single-layer dusty plasma suspension, particles
interact with a Yukawa potential~\cite{Konopka:2000} and are
damped by gas friction. Like 2D granular media~\cite{Rouyer:2000},
our particles are driven by an external source, and their average
kinetic energy is set by a balance of driving forces and
dissipation.

Random motion is characterized using the mean-square
displacement~(MSD) time series  and the probability distribution
function~(PDF). We define these separately for the $x$ and $y$
directions. For the $x$ direction, MSD is $\langle(x(\Delta
t)-x(0))^{2}\rangle$, and PDF is the histogram of $|x(\tau)-x(0)|$
evaluated at a specified delay $\tau$. Random motion characterized
as pure diffusion has two signatures: having a PDF that is
Gaussian at all times~\cite{Negrete:1998}, and obeying the
Einstein relation $\lim_{t\to\infty}\langle (x(\Delta
t)-x(0))^{2}\rangle=2D\Delta t$~\cite{Einstein:1905} at long
times, where $D$ is a diffusion constant.


Anomalous diffusion is a term used to describe motion that is
found empirically to not exhibit these signatures. In this case,
Fick's law is not satisfied, and there is no constant diffusion
coefficient.
Anomalous diffusion is associated with non-Gaussian statistics
sometimes, as in a liquid with vortices~\cite{Solomon:1993,
Ratynskaia:2006}, but not always~\cite{Liu:2007}.

Non-Gaussian statistics, which are often exhibited by
non-equilibrium systems, can follow a Tsallis distribution for a
random variable $z$, for example, displacement of a
particle~\cite{Tsallis:95}
\begin{equation}\label{Tsallis}
[1-\beta (1-q)z^{2}]^{1/(1-q)}
\end{equation}
where the non-extensivity parameter $q$ is $q=1$ for Gaussian and
$q\neq1$ for non-Gaussian statistics. In addition to anomalous
diffusion, Tsallis statistics has been used to describe turbulence
in pure electron plasmas~\cite{Boghosian:1996}.

There is a disagreement in the literature regarding whether motion
can ever be purely diffusive in 2D liquids. There are simulations
and theories predicting that motion is not diffusive for any type
of interparticle interaction~\cite{Alder:70,Ernst:1970}. On the
other hand, some recent simulations ~\cite{ Perera:98,
 Liu:2006} have found signatures of diffusion
in 2D liquids near the disordering transition, for various
interparticle interactions. All these simulations can have
limitations, including a finite system size.

The literature for diffusion in 2D liquids includes mostly
simulations of simple liquids. Experiments with 2D liquids are
fewer, including soft matter systems such as colloidal
suspensions~\cite{Marcus:1996}, granular
material~\cite{Wildman:1999}, and dusty
plasmas~\cite{Juan:1998,Juan:2001, Nunomura:2006,
Ratynskaia:2006}. Distinguishing diffusion from weakly anomalous
diffusion at a statistically significant level requires large data
sets for long times, which can be hard in a
simulation~\cite{Liu:2007} and even harder in an experiment due to
a limited thread lifetime. Here, the term ``thread" refers to a
time series of position measurements for one particle.

Two-dimensional transport has been experimentally studied in dusty
plasmas consisting of partially ionized rarefied gas that fills a
3D volume and contains charged microspheres suspended in a single
thin layer. Juan {\it et al}.~\cite{Juan:2001} measured the MSD
transverse to a sheared flow in a quasi-2D multilayer suspension,
and reported a diffusion coefficient that increased with the laser
power that generated the shear. Nunomura {\it et
al}.~\cite{Nunomura:2006} reported $D$ for a layer that was
perturbed by unstable movement of a few heavier particles in an
incomplete second layer thereby raising the suspension kinetic
temperature, $T$; they interpreted their results as indicating
diffusion, with $D\propto T$. Ratynskaia {\it et
al}.~\cite{Ratynskaia:2006} used a smaller cluster of microspheres
suspended in a single layer that was heated presumably by
instabilities in the dusty plasma; they described their motion as
superdiffusion, characterized by L\'evy statistics, driven by
flows~\cite{Ratynskaia:2006, Juan:2001}.  There have also been
other dusty plasma experiments to measure
shear~\cite{Nosenko:2004, Fortov:2005} and thermal
conduction~\cite{Nunomura:2005, Fortov:2007}. These experiments
can be compared with 2D granular medium experiments, which are
also driven-dissipative, and where diffusion takes a different
character if there is a flow~\cite{Utter:2004}.

Here we report an experimental study of 2D diffusion with a
different heating method. Our dusty plasma had no second
layer~\cite{Nunomura:2006}, and we used a laser heating configured
to avoid shear and macroscopic flows~\cite{Juan:2001,
Ratynskaia:2006}. The method provides steady conditions, allowing
us to record large data sets to provide a low detection limit for
superdiffusion.

Our laser-heated suspension is a driven-dissipative system.
Radiation pressure forces from rastered cw laser beams apply
strong kicks in the $\pm x$ directions to a few particles at a
time, and these collide with other particles to thermalize the
motion. Energy input from the laser is ultimately balanced by
dissipation on the neutral gas~\cite{Liu:2003}, allowing a
steady-state energy balance. Particles receive random kicks more
often from collisions with other particles than from the laser
beams.

We used a radio-frequency (rf) apparatus~\cite{Nosenko:2005},
Fig.~\ref{Setup}(a). Plasma was formed from Ar at 8.6 mTorr by
applying 13.56 MHz rf, yielding a dc self-bias of $-95$ V. We
introduced $>6000$ polymer microspheres, $4.83\pm0.08~\mu$m in
diameter, to form a single layer. The particles experienced a
damping rate $\nu_{E}=2.5$~s$^{-1}$~\cite{Liu:2003}. A cooled
camera recorded 37 s movies of particle motion at 55.56 frames per
second; its field-of-view (FOV) of $14.8 \times 10.8$ mm$^{2}$
included about 900 particles in the central portion of the
suspension. A 0.1 W Ar$^{+}$ laser beam swept horizontally once
per exposure illuminate particles. We verified that conditions
remained steady throughout the experiment, and using a side-view
camera we verified that no out-of-plane buckling occurred.

We analyzed movies, making $>2\times10^{7}$ particle positions
($x$, $y$) using the moment method, optimized as
in~\cite{Feng:2007}. Tracking particles from frame to frame yields
their trajectories, i.e., threads, Fig.~\ref{Setup}(b). Threads
have a finite half-life $\tau_{1/2}$, Fig.~\ref{Setup}(c), due to
particles exiting the FOV or becoming indistinguishable during
tracking. Lacking hydrodynamic coupling to any nearby wall, our
suspension rotated as a rigid body when it was in a solid state.
We always subtract this global rotation motion before calculating
other quantities. We report temperatures $T_{x}$ and $T_{y}$
computed from mean-square velocities.

The Wigner-Seitz radius~\cite{Kalman:2004} was $a=0.24$ mm. The
particle charge $Q/e=-5700$ and screening parameter
$a/\lambda_{D}=0.90$ were measured using the natural phonon
method~\cite{Nunomura:2006}. Repeated measurements of $Q$ and
$a/\lambda_{D}$ had a 4\% dispersion.  We calculated a time scale
~\cite{Kalman:2004} $\omega_{pd}^{-1}=9.2$ ms.

To provide a well-controlled and adjustable particle kinetic
temperature without affecting $Q$ or $a/\lambda_{D}$, we used the
laser-heating method of~\cite{Nosenko:2005}. A pair of 532 nm cw
laser beams directed in the $\pm x$ directions drew Lissajous
figures with frequencies $f_{x}=40.451$ Hz and $f_{y}=25$ Hz in a
16.2 mm wide rectangular stripe spanning the entire suspension in
the $x$ direction. We adjusted the temperature by varying the
laser power, Fig.~\ref{Setup}(c). Gradually increasing the power
of each beam to $P_{L}=0.75$ W, as measured in the chamber, we
melted the solid, as judged by $G_{\theta}$~\cite{Hartmann:2005}
and $g(r)$~\cite{Nosenko:2005}. Flow velocities were $<8\%$ of the
thermal velocity.


Compared to ~\cite{Nosenko:2005}, different Lissajous frequencies
and a wider stripe helped reduce coherent modes and
nonuniformities in the laser heating. Coherent modes, excited by
the finite frequency of the laser-beam's repetitive motion,
represented 10\% of the total power for the $x$ direction, and 6\%
for the $y$ direction. We measure $T$ and analyze random motion in
each of six portions of the FOV, i.e., cells, in
Fig.~\ref{Setup}(b) inset; this method reduces the effect of a
$35\%$ spatial variation of temperature over the full FOV. Due to
the anisotropic momentum input from our two laser beams, $T_{x}$
exceeded $T_{y}$, Fig.~\ref{Setup}(c), with $T_{x}/T_{y}$ as large
as 1.8.

The MSD, Fig.~\ref{MSD}, exhibits ballistic motion at short time,
$\Delta t<0.1$ s. For the $x$ direction, the MSD also exhibits an
artifact of laser heating between 0.1 and 1.0 s. Heating in the
$y$ direction arises only from collisions, so that it lacks this
artifact, leading us to analyze motion mostly in the $y$
direction. The MSD is always larger in the $x$ direction, due to
$T_{x}>T_{y}$. Here, we are interested in long-time behavior,
characterized by measuring the diffusion exponent
$\alpha$~\cite{Liu:2007}, which we present later.


The PDF has a self-similar form, as demonstrated in Fig.~\ref{PDF}
by the collapse of PDF curves at successive delays $\tau$ ranging
from 1 to 5 s. We scaled the displacement as $\Delta
\xi\equiv|y(\tau)-y(0)|/\tau^{\alpha/2}$ to yield a self-similar
form.

A Tsallis distribution  fits our self-similar PDF, and this fit
yields $q$. Recall that Gaussian statistics would be indicated by
$q=1$. For the lower laser power in Fig.~\ref{PDF}, we found
$q=1.08\pm0.01$; for the higher power we found $q =1.05\pm0.01$.
Repeating this analysis for other temperatures, and averaging the
resulting $q$ values within a temperature range, yields $\bar{q}$
in Table~\ref{tab:table1}.

We observe $\bar{q}>1$, i.e., non-Gaussian statistics for all
temperature ranges. These results are statistically significant,
with a confidence level $>95\%$ for rejecting the null hypothesis
$\bar{q}\leq1$. The deviation from Gaussian statistics is greatest
in our lowest temperature range, near the disordering transition.
Unlike~\cite{Ratynskaia:2006}, where a deviation from Gaussian
statistics was attributed to prominent vortex flows, our
experiment was prepared without flows.




By calculating the diffusion exponent $\alpha$, we determined that
random motion exhibits superdiffusion. In this test, we fit MSD
for displacements in the $y$ direction to the scaling $(\Delta
t)^{\alpha}$, yielding a measurement of $\alpha_{y}$ for different
ranges of $T_{y}$ and $\Delta t$. We then averaged the resulting
$\alpha_{y}$ values over ranges of $T_{y}$ and time delay. In
Table~\ref{tab:table1},  we report this mean value
$\bar{\alpha}_{y}$ and a standard deviation of the mean. In all
cases, we find $\bar{\alpha}_{y}>1$, i.e., superdiffusion, as
compared with $\alpha=1$ reported in \cite{Nunomura:2006}. The
level of superdiffusion, $\bar{\alpha}_{y}-1$, is small, but
statistically significant. Using our large data sets and the
student's t-test, we can reject the null hypothesis that
$\bar{\alpha}_{y} \le 1$ with a confidence level of $>95$\% (i.e.,
$p < 0.05$) in all cases except possibly the highest temperature
range in Table~\ref{tab:table1}.

The level of superdiffusion we detected did not reach the larger
values of $\alpha > 1.1$ observed in equilibrium 2D Yukawa
simulations~\cite{Liu:2007}. Like our experiment, these
simulations have no flows and no incomplete second layer, but they
differ by assuming an energy equilibrium with no friction. We
speculate that friction may suppress superdiffusion. Experimental
systems experience friction, more strongly for colloids which are
overdamped, and less strongly for dusty plasma, which uses a
rarefied gas rather than a liquid solvent.

\begin{table}
\caption{\label{tab:table1}The measure $\bar{q}$ of
non-extensivity indicates non-Gaussian statistics. Mean diffusion
exponent $\bar{\alpha}_{y}$ (and p-value for testing the null
hypothesis that there is no superdiffusion) indicates
superdiffusion.}
\begin{ruledtabular}
\begin{tabular}{lllll}
&    time &  \multicolumn{3}{c}{temperature range $T_y$
($10^{3}$~K)}\\ & delay (s)  & $10-20$ & $20-40$ & $40-60$\\
\hline $\bar{q}$&
$1<\tau<5$&$1.20\pm0.09$&$1.11\pm0.06$&$1.05\pm0.02$\\ \hline
$\bar{\alpha}_{y}$&$1<\Delta t
<5$&$1.052\pm0.019$&$1.059\pm0.011$&$1.009\pm0.011$\\ p && 0.007 &
$2\times10^{-5}$ & 0.210  \\ $\bar{\alpha}_{y}$& $5<\Delta t
<9$&$1.088\pm0.048$&$1.082\pm0.040$&\\ p& & 0.042 & 0.038
\\ $\bar{\alpha}_{y}$& $9<\Delta t<13$&$1.146\pm0.062$\\ p&&
0.028\\ $\bar{\alpha}_{y}$&$13<\Delta t<17$&$1.183\pm0.064$\\ p&&
0.006\\
\end{tabular}
\end{ruledtabular}
\end{table}

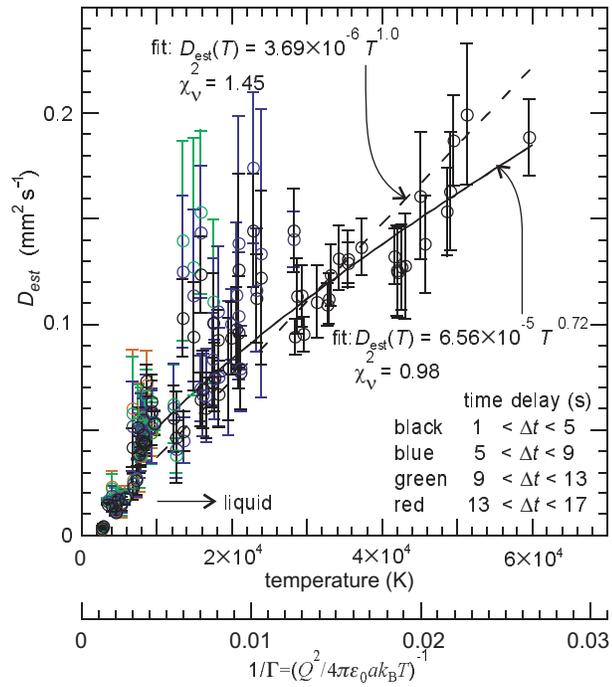
\begin{figure}[p]
\caption{\label{DvsT} (Color online) Temperature dependence of
$D_{est}$ calculated from $\langle(y(\Delta
t)-y(0))^{2}\rangle/2\Delta t$. The smooth curve is a fit,
weighted by error bars, of data for a liquid. Scatter is due to
finite time interval and number of threads.}
\end{figure}

Finally, we use our large data sets to investigate whether the MSD
in our 2D liquid can yield a diffusion coefficient that has the
scaling with temperature previously reported for an
experiment~\cite{Nunomura:2006}. We tentatively calculate an
estimated diffusion coefficient $D_{est}$ as $\langle(y(\Delta
t)-y(0))^{2}\rangle/2\Delta t$ for displacement in the $y$
direction, averaged for all movies and for various time delays
$\Delta t$, yielding Fig.~\ref{DvsT}. This coefficient of course
increases with temperature. Based on a smaller data set, it was
reported~\cite{Nunomura:2006} that in a liquid this increase fits
a straight line with an intercept on the horizontal axis at the
melting temperature. However, we found that such a line fits our
data more poorly ($\chi^{2}_{\nu}>4$) than a power law
($\chi^{2}_{\nu}=0.98$) or a straight line through the origin
($\chi^{2}_{\nu}=1.45$).

We thank Z. Donk\'o, V. Nosenko, J. Scudder, and F. Skiff for
helpful discussions. This work was supported by NASA and DOE.

\begin{figure}[p]
\caption{\label{Setup}(a) Apparatus. An argon plasma is formed in
a vacuum chamber, not shown here. Microspheres suspended in a
single layer above a lower electrode are kicked by a pair of
rastered laser beams. (b) Particle trajectories for $P_{L} =2.5$~W
in cell 1. We divide the camera's full field-of-view (FOV) into 6
cells, as shown in the inset. Motion is mostly ballistic in the
0.5 s time interval shown here. (c) Heating. Kinetic temperatures,
averaged over the full FOV, increase with $P_{L}$. Structures with
$G_{\theta}<0.1$ were identified as liquid and lacked large
crystalline domains; structures identified as solid were not
defect-free. At higher temperatures, threads have a shorter
half-life $\tau_{1/2}$. The anisotropy $T_{x}>T_{y}$ is due to
laser heating.}
\end{figure}

\begin{figure}[p]
\caption{\label{MSD} (Color online) MSD, averaged over threads and
overlapping time segments~\cite{Saxton:1997}, computed separately
for $x$ and $y$. Data here are for two laser powers. The MSD in
the $x$ direction is larger than that in the $y$ direction, due to
$T_{x}>T_{y}$. Distance is normalized by the Wigner-Seitz radius
$a$.}
\end{figure}


\begin{figure}[p]
\caption{\label{PDF} (Color online) PDF for the $y$ direction, and
the same conditions as in Fig.~\ref{MSD}. Self-similarity is
revealed by the collapse of PDFs at successive delays $\tau$, with
the scaled displacement $\Delta
\xi\equiv|y(\tau)-y(0)|/\tau^{\alpha/2}$. Fitting to
Eq.~(\ref{Tsallis}) yields the smooth solid curves and the measure
of non-extensivity $q$. Other curves shown bracket the smooth
curve for the fit. Data here are for $1\leq\tau\leq5$~s.}
\end{figure}

\end{document}